# Epigenetic Landscape of Interacting Cells: A Model Simulation for Developmental Process

**Motoki Nakagawa・Osamu Narikiyo**

*Department of Physics, Kyushu University, Fukuoka 810-8560, Japan*

**Abstract**

We propose a physical model for developmental process at cellular level to discuss the mechanism of epigenetic landscape. In our simplified model, a minimal model, the network of the interaction among cells generates the landscape epigenetically and the differentiation in developmental process is understood as a self-organization. The effect of the regulation by gene expression which is a key ingredient in development is renormalized into the interaction and the environment. At earlier stage of the development the energy landscape of the model is rugged with small amplitude. The state of cells in such a landscape is susceptible to fluctuations and not uniquely determined. These cells are regarded as stem cells. At later stage of the development the landscape has a funnel-like structure corresponding to the canalization in differentiation. The rewinding or stability of the differentiation is also demonstrated by substituting test cells into the time sequence of the model development.

# 1 Introduction

The scenario of epigenetic landscape [1,2], whose illustration is shown in Fig. 1, is a powerful concept for describing developmental phenomena. It is applied to wide class of phenomena from molecular to sociological levels [3-9]. However, the physical process generating the landscape has not been explicitly demonstrated so far. Thus in this paper we propose a physical model for developmental process at cellular level and explain how the epigenetic landscape arises.

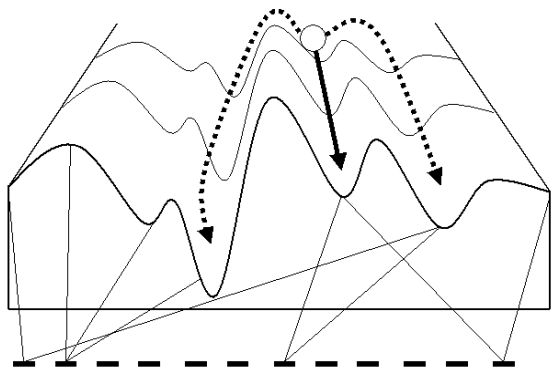

**Fig. 1** An epigenetic landscape of a developmental process. The arrows show the direction of time and represent possible processes of the development. In a cross section the state of the developing system at a time is placed in the horizontal direction and the fitness of the state in the vertical direction where the lower value corresponds to the higher fitness. In ordinary explanation the landscape is determined by the regulations by gene expression schematically shown by bars and lines outside the landscape where the bars represent genes and the lines regulations.

Although the regulation by gene expression, for example, via DNA methylation or via morphogen, is a key ingredient in the development [10,11] at cellular level, we focus our attention onto the interaction among cells as the driving force to generate the epigenetic landscape. In our modeling the effect of the gene expression is renormalized into the interaction and the environment of the cells and considered implicitly. The importance of the interaction is emphasized in the study of dynamical systems approach [12,13] and it is also the heart of our study.

This paper is organized as follows. In Section 2 we introduce our model and simulation rule. Various simulation results are discussed in Section 3. Our conclusion is given in Section 4.

## 2 Model

2-1 Fitness

Here we propose a physical model for an epigenetic landscape in developmental process at cellular level. The epigenetic landscape of our model is a time-dependent fitness landscape for the total system consisting of a group of interacting cells and its environment.

In our simplified model, a minimal model, the time-dependent energy landscape determined by the interaction among cells is identified with the epigenetic fitness landscape.

We consider the case where each cell has only two kinds of chemicals, c1 and c2, as its constituent. The number of c1 and c2 in the $i$-th cell are represented by the variables, $s_i^{(1)}$ and $s_i^{(2)}$, respectively. The total number of chemicals in a cell is fixed as $s_i^{(1)} + s_i^{(2)} = s_0$ with $s_0$ being a constant.

The chemical reactions in the $i$-th cell and the ones in its neighboring cells are mutually influenced and form a network. The fitness of the $i$-th cell, $F_i$, is determined by neighboring cells as

$$F_i = J_{i,0}^{(1)} s_i^{(1)} + J_{i,0}^{(2)} s_i^{(2)} + \sum_{k=1}^{3} [J_{i,k}^{(1)} (s_{i-k}^{(1)} + s_{i+k}^{(1)}) + J_{i,k}^{(2)} (s_{i-k}^{(2)} + s_{i+k}^{(2)})], \qquad (1)$$

where the couplings $J_{i,k}^{(1)}$ and $J_{i,k}^{(2)}$ represent the influences of the $k$-th neighbors via the chemical reaction network. Here $J_{i,k}^{(1)}$ and $J_{i,k}^{(2)}$ are not constants but change according to the feedback represented by $J_{i,0}^{(p)} = J_0^{(p)} s_i^{(p)}$ and $J_{i,k}^{(p)} = J_k^{(p)} (s_{i-k}^{(p)} + s_{i+k}^{(p)})$ where each $J_k^{(p)}$ is a constant and $p = 1$ or $p = 2$. Such a non-linear feedback imitates hypercycle-type interaction [14] and takes the effect of the regulation by gene expression into account.

In the following we set all the couplings positive so that no frustration is introduced into this model. Such frustration free couplings lead to a funnel structure in the fitness landscape as discussed later. If both positive and negative couplings are employed, a frustration is introduced and resulting landscape may become random as glass or spin-glass.

With the fitness $F_i$ for each cell we represent the fitness of the total system by the energy $E$ given by

$$E = -\sum_{i=1}^{N} F_i, \qquad (2)$$

where $N$ is the total number of cells in the system. In this representation the state with lower values of $E$ is favored and $E$ plays the same role as the potential energy in physics. By the definition $E$ corresponds to the mean of the individual fitness $F_i$.

In our model the couplings $J_{i,k}^{(1)}$ and $J_{i,k}^{(2)}$ among cells are the driving force to generate the epigenetic landscape. Moreover we explicitly construct the environment for the cells as discussed later and its development reflects the precedent state of the cells. The effect of the regulation by gene expression, for example, via DNA methylation or via morphogen, which is a key ingredient in development [10,11], is renormalized into the interaction and the environment.

The fitness $F_i$ accounts for the fact that the state of a cell is regulated by the chemical reaction network in which it is embedded. The environment of each cell is altered according as the cell development. On the other hand, this alteration regulates cells. Namely, each cell and its environment co-evolve.

Since we would like to simulate cell development against fluctuation and the implementation of fluctuation is easy on the basis of the above-mentioned model with the Monte-Carlo simulation as discussed in the following, we choose this model. Recent experiments have uncovered the fact that a living cell feels large fluctuations even for the gene expression [15,16].

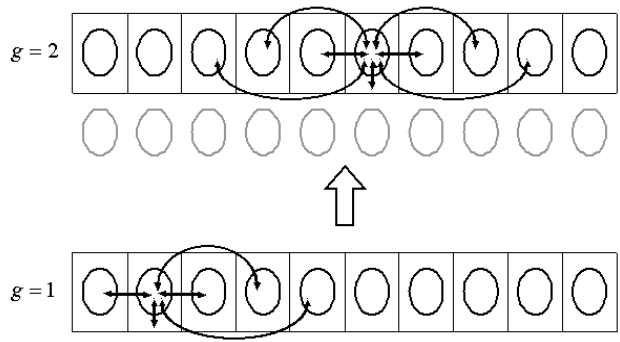

**Fig. 2** Schematic representation of the simulation. Each cell is depicted by an oval and put in a square representing its environment. In the first generation, $g=1$, the interactions of a selected cell ($i=2$) are shown by lines with arrows. In the second generation, $g=2$, the interactions of a selected cell ($i=6$) are shown and the states of cells outside of squares are frozen.

2-2 Monte-Carlo Simulation

Using above-mentioned model for the fitness we perform the Monte-Carlo simulation of developing cells in the following procedure.
(1) As the first generation we prepare $N$ cells forming a line where each cell has $s_0$ chemicals and is put inside its own environment. The environment for the $i$-th cell also consists of two kinds of chemicals, c1 and c2, whose numbers, $e_i^{(1)}$ and $e_i^{(2)}$, are under the condition $e_i^{(1)} + e_i^{(2)} = e_0$ with $e_0$ being a constant. For convenience each environment is expressed as a box and $N$ boxes also form a line as shown in Fig. 2. All the numbers, $s_i^{(1)}$, $s_i^{(2)}$, $e_i^{(1)}$ and $e_i^{(2)}$, are chosen randomly. Then we calculate the total fitness $E$ of the initial state.
(2) We perform one Monte-Carlo step procedure explained later.
(3) As the next generation we make a copy of each cell. The copy is as the same as its original cell and inherits the original's environment. After making copies the original cells are put outside the environment and frozen. Then the copies are put inside the corresponding environment.
(4) We perform one Monte-Carlo step procedure for the copy where the copy plays the same role as the former original. For simplicity the interaction between generations is neglected.
(5) We repeat the procedure (3) and (4).

Within one Monte-Carlo step procedure each cell experiences $e_0$ times of elementary processes on average. The elementary processes consist of two kinds of trials. One is the randomly chosen exchange of chemicals between two neighboring cells. The other is the randomly chosen exchange between a cell and its environment. The trial is accepted or rejected according to the Metropolis rule: it is always accepted if the exchange leads to the decrease of $E$, but it is accepted by the probability

$\exp(-\Delta E/T)$ if the exchange leads to the increase of $E$ by $\Delta E$. Even if the trial fails, it is counted as experience. Here we have introduced a parameter $T$, which corresponds to the temperature in physics, in order to take fluctuation into account in a phenomenological manner. Reflecting the fact that each cell loses the susceptibility, for example by gene-expression regulation, to its neighboring cells and environment with the passage of time, we scale the parameter in the $g$–th generation as $T = T_0/g$ with $T_0$ being a constant.

## 3 Results and Discussions

### 3-1 Canalization

We have performed the Monte-Carlo simulation described above with the interaction parameters $J_0^{(1)}/J_0^{(2)} = J_1^{(1)}/J_1^{(2)} = J_2^{(1)}/J_2^{(2)} = J_3^{(1)}/J_3^{(2)} = 1.25$ and $J_0^{(2)} = 10$, $J_0^{(2)} = 5$, $J_0^{(2)} = 3$, $J_0^{(2)} = 1$. The numbers are fixed as $N = 10$, $s_0 = 20$ and $e_0 = 20$. The simulation time is measured by the number of exchange of chemicals and the system experiences $Ne_0$ exchanges in a generation.

In the case of relatively lower temperature $T$, namely smaller fluctuation, the system smoothly reaches its lowest energy state, namely its highest fitness state, as shown in Fig. 3. In the lowest energy state adjacent cells exploit different chemicals to avoid competition so that $s_i^{(1)} = 0$ or $s_i^{(2)} = 0$ in the differentiated state.

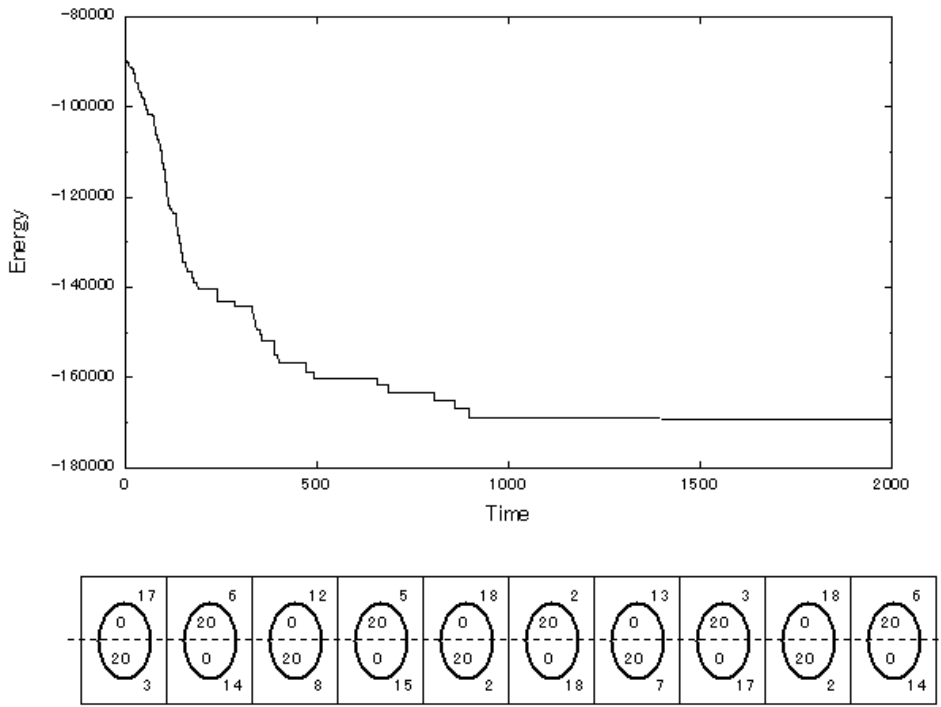

**Fig. 3** Time dependence of the energy $E$ for $T_0 = 10$. The numbers of the chemicals are shown,

for example, as $s_4^{(1)} = 20$, $s_4^{(2)} = 0$, $e_4^{(1)} = 5$ and $e_4^{(2)} = 15$.

In our simulation the temperature $T$ becomes smaller at later generations by the scaling procedure described above. This scaling, effectively representing the fact that each cell loses the susceptibility to its neighboring cells and environment with the passage of time, stabilizes the system even if the temperature is relatively high and the system fluctuates without it as shown in Fig. 4.

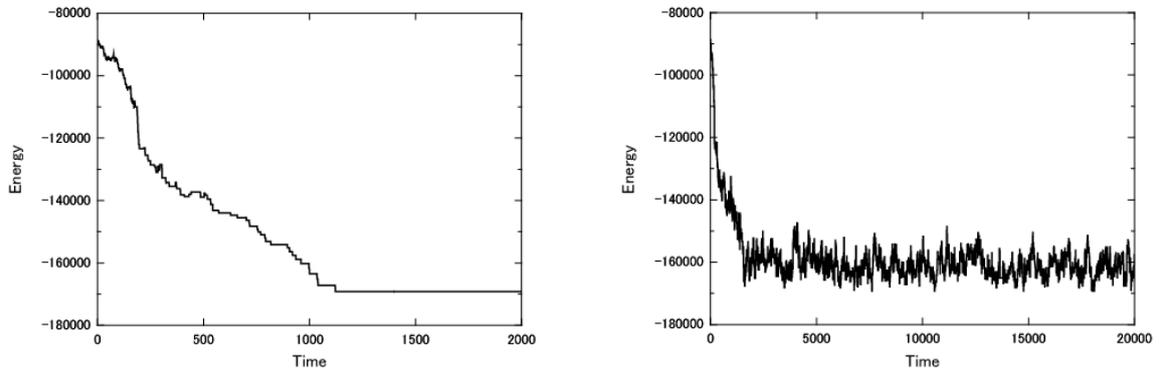

**Fig. 4** Time dependence of the energy $E$ for $T_0 = 1000$. **a** Left: the temperature is scaled as $T = T_0 / g$, while **b** Right: it is fixed as $T = T_0$.

To know the energy landscape around which the system is located we plot the energy of nearby states as show in Fig. 5. In Fig. 5a the energy of the states generated from the initial state by exchanging chemicals between cells and their environments are shown. These states are classified by the Hamming distance from the initial state where a single exchange of c1 and c2 leads to 2 Hamming distance. The energy level of the initial state with random composition of chemicals is located in the middle of crowded energy-level distribution. Namely, at earlier stage of the development the energy landscape is rugged with small amplitude. The state of cells in such a landscape is susceptible to fluctuations and not uniquely determined. These cells are regarded as stem cells which are plastic and not differentiated.

In Fig. 5b the energy of the states generated from the lowest energy state by exchanging chemicals between cells and their environments are shown. The energy levels of generated states are separated by large gap from the lowest energy level. Namely, at later stage of the development the energy landscape has a funnel-like structure corresponding to the canalization in differentiation. Such a funnel structure is similar to the case of protein folding [17].

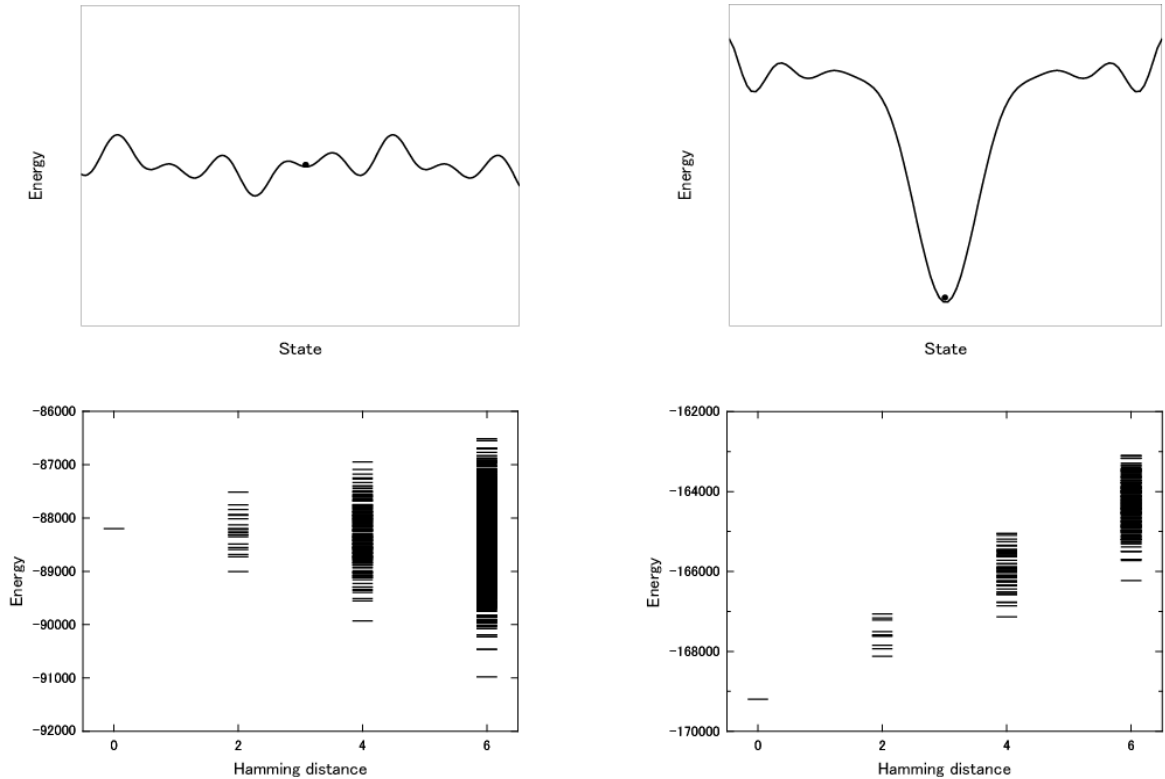

**Fig. 5** **a** Left: energy landscape around the initial state and **b** Right: energy landscape around the ground state for the simulation in Fig. 4a. Upper panels show the schematic representation of the landscape. Lower panels show the energy level distribution.

We have observed a self-organization, differentiation, of the system by interactions among cells representing the chemical reaction network. In the course of differentiation each cell chooses its own chemical reaction to avoid competition with the other cells and the system accomplishes division of labor. In the chemical reaction network the state of a cell is not determined individually but cooperatively under the influence of the other cells. To stabilize the differentiated state the system needs to reduce the susceptibility to fluctuation. The resulting epigenetic landscape has a funnel structure corresponding to the canalization in differentiation.

3-2 Context

The behavior of each cell is determined by interaction with surrounding cells and depends on the context in which it is embedded. We examine the rewinding or stability of the differentiation by substituting test cells into the system.

In Fig. 6 the time dependence of the test cell substituted into an undifferentiated system is shown. In spite of that the test cell is set to have the same chemical components as the differentiated cell, it behaves in harmony with surrounding undifferentiated cells so that the cell state is rewound to undifferentiated one. The test cell feels a rugged energy landscape and behaves as a stem cell which is susceptible to fluctuation and open to diversity of chemical component.

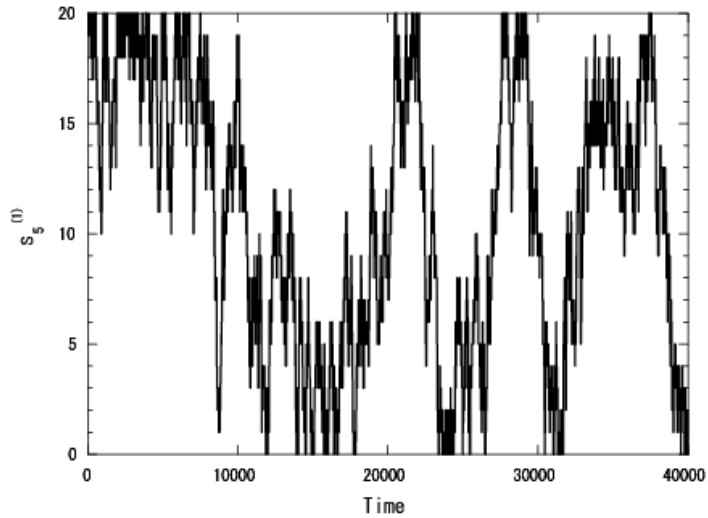

**Fig. 6** Time dependence of the number of a chemical $s_5^{(1)}$ in the test cell for $T_0 = 10000$. At time $= 0$ an undifferentiated cell ($i = 5$) in the random initial state is replaced by the differentiated test cell with $s_5^{(1)} = 20$ and $s_5^{(2)} = 0$.

On the other hand, a few undifferentiated cells follow the surrounding differentiated cells into which they are substituted as shown in Fig. 7. Thus the stability of the differentiated system against perturbation has been confirmed. This simulates the fact that stem cells are induced by the surrounding cells to acquire the same function as the surrounding.

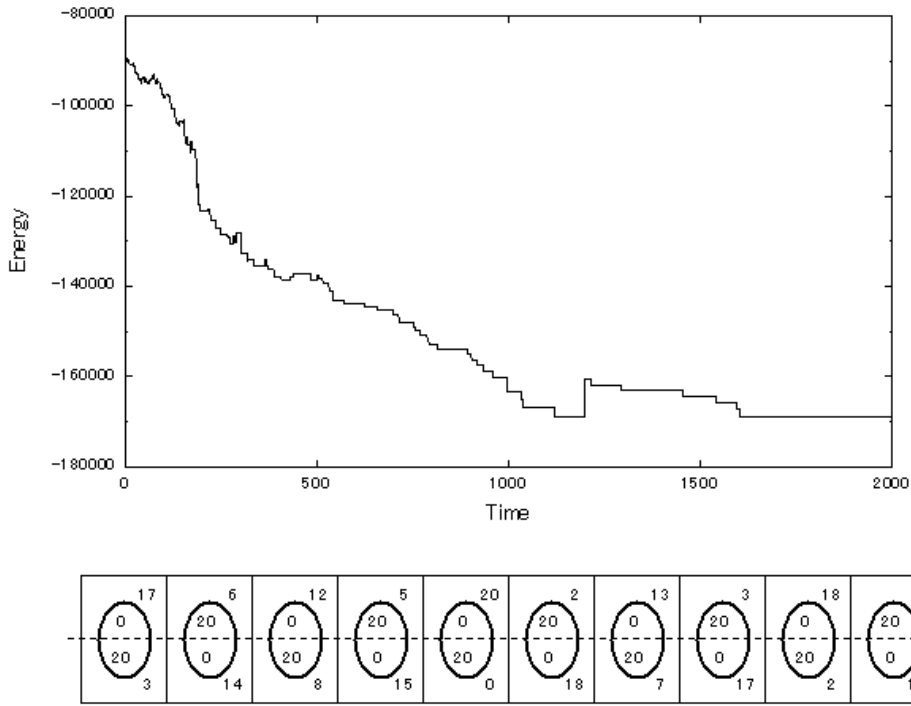

**Fig. 7** Time dependence of the energy $E$ for $T_0 = 1000$. The numbers of the chemicals are shown, for example, as $s_4^{(1)} = 20$, $s_4^{(2)} = 0$, $e_4^{(1)} = 5$ and $e_4^{(2)} = 15$. At time $= 1200$ an differentiated cell ($i = 5$) in the ground state is replaced by a undifferentiated random cell.

When the substituted cells are not regarded as a minority, the total system settles down to a new niche as shown in Fig. 8. In this case the perturbation by substituted cells exceeds the restoring force of the original system and destroy the function of the differentiated cells. This simulates the fact that the stable system of differentiated cells has the limitation of its tolerance.

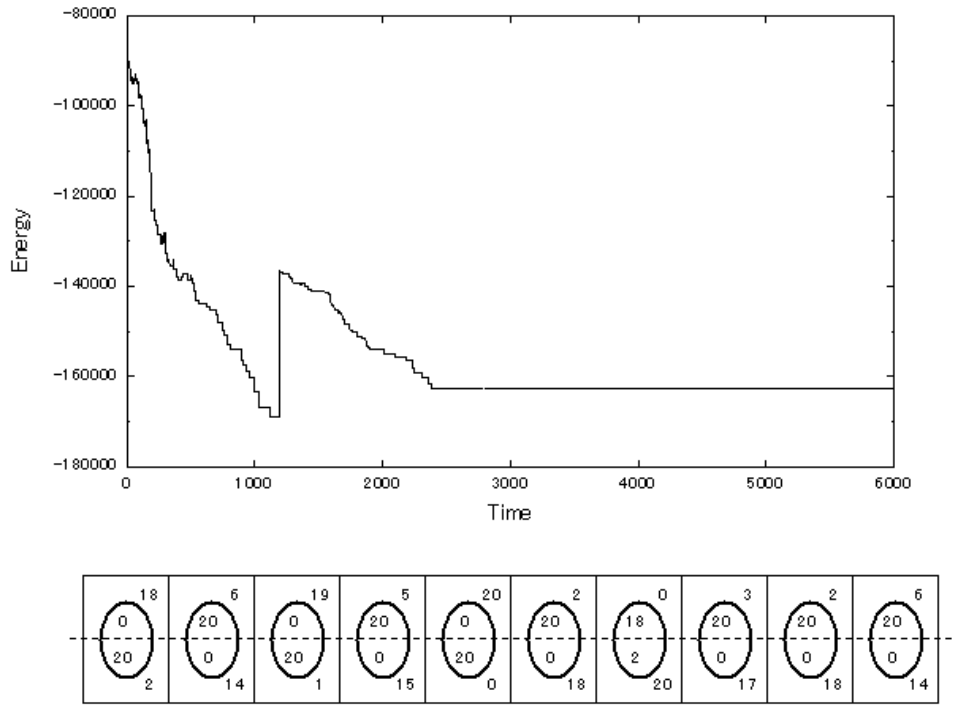

**Fig. 8** Time dependence of the energy $E$ for $T_0 = 1000$. The numbers of the chemicals are shown, for example, as $s_4^{(1)} = 20$, $s_4^{(2)} = 0$, $e_4^{(1)} = 5$ and $e_4^{(2)} = 15$. At time $= 1200$ differentiated cells ($i = 1,3,5,7,9$) in the ground state are replaced by undifferentiated random cells.

## 4 Conclusion

We have proposed a simple model, a minimal model, to discuss physical mechanism of epigenetic landscape in developmental process at cellular level. By the Monte-Carlo simulation the canalization of differentiation has been demonstrated. In our model the interaction among cells leads to the differentiation in developmental process as a self-organization.  The effects of gene are taken into account only implicitly, while the initial state of developmental process is assigned by the information of gene and the interaction among cells and the susceptibility to fluctuation are regulated by gene expression.

It has been also shown that the behavior of each cell is determined by interaction with surrounding cells and depends on the context in which it is embedded. The rewinding or stability of differentiation has been demonstrated by substituting test cells into the system.

Since the aim of our paper is to demonstrate the importance of the interaction by a simulation on the basis of a minimal model, our model is too simple and abstract. Although we should employ more realistic model in order to describe biological phenomena, such elaboration is left to future study.

**Appendix   STAP cells**

Recently reported context-dependent behavior of STAP cells [Nature **505**, 641-647 (2014)] can be regarded as an example of that in our model as discussed in the section 3-2 of the main text. It should be noted that such a behavior is the result of the interaction between the cell and its environment, while the behavior of iPS cells is the consequence of the regulation by genes.